\documentclass[twocolumn,trackchanges]{aastex631}
\newcommand{\bprpDered}{$(G_{\rm BP}-G_{\rm RP})_0$}
\newcommand{\absGDered}{$(M_G)_0$}

\begin{document}

\title{They Won't Be Giants: Missing Metal-Rich RGB Stars in Gaia Data Indicate Truncated Stellar Evolution}

\newcommand{\msun}{$M_{\odot}$}
\newcommand{\osu}{Department of Astronomy, The Ohio State University, Columbus, 140 W 18th Ave, OH 43210, USA}
\newcommand{\ccapp}{Center for Cosmology and Astroparticle Physics (CCAPP), The Ohio State University, 191 W. Woodruff Ave., Columbus, OH 43210, USA}

\author[0000-0003-4769-3273]{Yuxi(Lucy) Lu}
\affiliation{\osu}
\affiliation{\ccapp}

%\author[0009-0001-1470-8400]{K. Z. Stanek}
%\affiliation{\osu}
%\affiliation{\ccapp}

\author[0000-0003-0929-6541]{Madeline Howell}
\affiliation{\osu}
\affiliation{\ccapp}

\author[0000-0002-7549-7766]{Marc H. Pinsonneault}
\affiliation{\osu}
\affiliation{\ccapp}

\author[0000-0003-0174-0564]{Andrew R.\ Casey}
\affiliation{Center for Computational Astrophysics, Flatiron Institute, 162~Fifth~Ave., New~York, NY~10010,~USA}
\affiliation{School of Physics and Astronomy, Monash University VIC 3800, Australia}

\author{Jos\'e G. Fern\'andez-Trincado}
\affiliation{Centro de investigaci\'on en Astronom\'ia, Facultad de Ingenier\'ia, Ciencia y Tecnolog\'ia, Universidad Bernardo O’Higgins, Av. Viel 1497, Santiago, 8370993, Chile}

\author[0000-0002-6972-6411]{Jos\'e Eduardo M\'endez Delgado}
\affiliation{nstituto de Astronomıa, Universidad Nacional Aut\'onoma de M\'exico, A.P. 70-264, 04510, Mexico, D.F., Me\'xico}



\begin{abstract}
We investigate the population of luminous red giant branch stars as a function of metallicity using Gaia XP metallicity combined with SDSS-V, GALAH, and LAMOST. 
After applying uniform selection criteria and extinction corrections, we construct absolute magnitude distributions across metallicity bins spanning [Fe/H] $=-1$ to $>0.4$. 
We find a systematic deficit of luminous giants at high metallicity, while the red clump and lower red giant branch populations remain largely unchanged. 
This behavior is consistent with enhanced mass loss at high metallicity, arising from either binary interactions or single-star evolution.
This trend is robust across multiple surveys and persists within volume-limited subsamples (1-4 kpc), suggesting it is not driven by distance or selection effects.
Synthetic stellar populations based on PARSEC isochrones reproduce the overall magnitude distributions but do not predict a decline in luminous giants with metallicity. 
Tests of potential systematics, including extinction effects and metallicity scale consistency using open clusters, do not account for the observed trend. 
We also find no evidence that survey-to-survey differences in metallicity drive the observed result. 
Together, these findings suggest a metallicity-dependent reduction in the number of luminous red giants that is not captured by current models.
This result may have implications for stellar evolution at high metallicity, helium white dwarf formation, and the initial mass function as well as the UV upturn in metal-rich galaxies.

\end{abstract}



\section{Introduction} \label{sec:intro}
Mass loss during the red giant branch (RGB) phase is a fundamental yet poorly understood process that shapes the late-stage evolution of low- and intermediate-mass stars.
By removing the hydrogen-rich envelope before or during the helium flash, RGB mass loss determines the subsequent evolutionary pathways of stars, influencing the populations of horizontal branch (HB) stars, subdwarf stars, and white dwarfs (WDs). 
Despite its importance, the physical mechanisms governing RGB mass loss, particularly at high metallicity, remain uncertain.

Traditional theoretical prescriptions for RGB mass loss \citep[e.g.,][]{Reimers1975, Schroder2005} predict that the mass-loss rate scales with ($LR/M$), where $L$, $R$, and $M$ are the stellar luminosity, radius, and mass, respectively. 
Although these prescriptions do not include an explicit metallicity dependence, their application to stellar models generally leads to higher mass-loss rates at higher metallicity, as metal-rich stars have increased envelope opacity and larger radii at fixed mass.
While integrated RGB mass loss inferred from globular cluster populations appears to increase with metallicity over the range of [Fe/H] $\lesssim -0.5$ \citep[e.g.,][]{McDonald2015, Tailo2020, Howell2022, Howell2024, Howell2025}, asteroseismic measurements suggest the opposite trend at higher metallicities. 
Analyses of both open clusters and field stars indicate that RGB mass loss likely decreases with increasing metallicity over the range of $-0.5 < {\rm [Fe/H]} < 0.4$ \citep{Miglio2012, Li2025, Roberts2026, pinsonneault2024, Ash2025, Howell2026}.
Since the metallicity dependence of RGB mass loss may not be monotonic, it is possible that the trend observed at intermediate metallicities could reverse again at the extreme metal-rich end ([Fe/H]$>0.4$). 
However, such stars are rare in the solar neighborhood, likely because they preferentially formed in the inner Galaxy, where the metallicity is higher, and subsequently migrated outward \citep[e.g.,][]{Sellwood2002, Lu2024_migration}.

Extreme RGB mass loss at the highest metallicities could have broader implications for stellar population studies. 
One possible consequence is the production of hot, stripped stellar remnants that contribute to the ultraviolet (UV) upturn observed in quiescent early-type galaxies, where an excess of flux at ($\lambda < 3000\AA$) is detected beyond that expected from their old, metal-rich stellar populations \citep[e.g.,][]{Code1979}. 
The strength of the UV upturn has also been observed to increase with metallicity \citep[e.g.,][]{Akhil2024}, suggesting that metallicity-dependent envelope stripping may play a role. 
In addition, preferential removal of metal-rich RGB stars through enhanced mass loss could affect the integrated light of massive elliptical galaxies and potentially contribute to the apparent bottom-heavy stellar initial mass functions (IMFs) inferred from their spectra \citep[e.g.,][]{Conroy2012, Cheng2026}. 
In this scenario, the inferred excess of low-mass stars could instead arise, at least in part, from a deficit of metal-rich giant stars rather than an intrinsically bottom-heavy IMF.

Directly detecting the descendants of extreme RGB mass loss, such as subdwarf stars, offers a potential way to constrain whether envelope stripping becomes more efficient at high metallicity. 
However, this approach is observationally challenging. 
The stripped remnants are typically hot, faint, and short-lived, making them difficult to identify in the field sample, particularly when they originate from rare metal-rich populations. 
Furthermore, distinguishing genuine post-RGB stripped stars from other hot stellar populations requires precise stellar parameters and evolutionary modeling. 
Consequently, the absence or presence of detected subdwarfs in metal-rich environments remains difficult to interpret as a direct constraint on RGB mass-loss efficiency.

The observational challenges of identifying stripped remnants are particularly relevant for old, metal-rich environments, where direct constraints on extreme RGB mass-loss channels remain limited.
The old, metal-rich open cluster NGC 6791 provides a unique laboratory for studying such processes, hosting a large population of low-mass helium white dwarfs (HeWDs; $\lesssim 0.45,M_{\odot}$; \citealt{Liebert2005, Kepler2007}).
The existence of HeWDs poses a challenge to standard single-star evolution models.
The evolution of low-mass stars that fail to ignite helium and directly form HeWDs would require longer than a Hubble time.
As a result, the large HeWD population in NGC 6791 is generally thought to form primarily through binary interactions, in which the envelope of a RGB star is stripped by a companion before the helium flash, leaving behind a low-mass HeWD core \citep[e.g.,][]{Nelemans1998, Justham2009, Geier2015}.
This scenario is consistent with the large population of extreme horizontal branch (HB) stars in the cluster, which are hotter and bluer HB stars with thin hydrogen envelopes that can also be produced through binary interactions.
In addition, the high binary fraction observed among field HeWDs further supports this formation pathway.
The binary fraction reaches nearly $\sim$100\% at $\sim$0.2,$M_{\odot}$ and decreases to $\sim$70\% at $\sim$0.4,$M_{\odot}$ \citep[e.g.,][]{Brown2010, Brown2011}.

It is worth noting, however, that not all HeWDs have detectable companions, even after accounting for orbital inclination effects \citep[e.g.,][]{Brown2011, Munday2024}.
If some HeWDs are truly single, additional evolutionary channels may be required.
For example, two HeWDs may merge without igniting helium if their total mass remains below the helium-flash threshold ($\lesssim 0.45,M_{\odot}$).
However, most WD mergers are expected to involve more massive remnants, producing either carbon/oxygen WDs or subdwarf B stars, which are helium-core-burning stars with thin hydrogen envelopes \citep[e.g.,][]{Saio2000, Han2002, Nelemans2010}.
Another proposed channel involves the merger of a low-mass main-sequence star ($\lesssim 0.6,M_{\odot}$) with a HeWD, forming an RGB-like object that resembles a more evolved low-mass star.
The remnant subsequently evolves into a HeWD \citep[e.g.,][]{Zorotovic2017, Zhang2018}.

In the context of NGC 6791, these results suggest that subgiant stars may follow two distinct evolutionary pathways.
In the first channel, stars evolve normally onto the RGB, retain typical stellar masses, and appear in asteroseismic samples, thereby biasing the inferred integrated mass loss toward low values.
In the second channel, stars belonging to a potential putative ``missing RGB population'' lose their envelopes and evolve away from the giant branch to become subdwarfs and eventually HeWDs, either through binary interactions or enhanced mass loss.
The channel responsible for producing HeWDs may become increasingly important at high metallicity, as most clearly illustrated by NGC 6791, while the surviving RGB population continues to exhibit only modest integrated mass loss.
If this interpretation is correct, it would imply a substantial reduction in the number of RGB stars at super-solar metallicities.

In this paper, we report a deficit of extremely metal-rich giant stars ([Fe/H] $> 0.4$) across four large spectroscopic surveys, a result that cannot be explained solely by population age effects.
The survey selection and synthetic populations are described in \autoref{sec:datamethod}.
The missing giants are presented in \autoref{sec:result}, while comparisons with synthetic populations are discussed in \autoref{subsec:synth} and \autoref{subsec:ratio}.
We further explore several possible explanations, including observational biases against detecting metal-rich stars (\autoref{subsec:bias}), and potential misclassification of metal-rich stars in survey pipelines (\autoref{subsec:misclass}).

\section{Data \& Methods} \label{sec:datamethod}
\subsection{Photometric and Metallicity Data}\label{subsec:data}
We obtain photometry data such as the $G$, $G_{\rm BP}$, $G_{\rm RP}$ band magnitudes and the parallax data from Gaia Data Release 3 (DR3). 
We selected stars that are within 4 kpc of the Sun and with parallax-over-error $>$ 5 to ensure we have relatively accurate photometry.
We adopt a distance limit of 4 kpc, as Gaia XP spectra provide the most accurate metallicity measurements down to $G = 16$, corresponding to an absolute magnitude of $\sim 3$ and covering the majority of giant stars.

The bulk metallicity ([M/H]) or iron abundances ([Fe/H]) are obtained from large scale spectroscopic survey catalogs.
They include [M/H] measurements from Gaia XP spectra, determined by \cite[][]{Andrae2023}, who trained the machine learning algorithm XGBoost \citep{Chen2016} on APOGEE DR17 stellar parameters \citep{Abdurrouf2022}; [M/H] measurements from SDSS-V APOGEE Internal Data Product 4 \citep[IPL4,][]{Kollmeier2017, Wilson2019, Kollmeier2026}; [Fe/H] measurements from GALAH Data Release 4 \citep[DR4,][]{Buder2025}, and [Fe/H] measurements from LAMOST Data Release 11 \citep{Cui2012}.
The Gaia XP metallicity provides the largest dataset there is for metallicity of field stars, and the SDSS-V metallicities provide a large sample of stars that have metallicity determined from traditional methods and high resolution spectra.
Lastly, GALAH DR4 and LAMOST DR11 serve as independent validations of our results as the metallicity from Gaia XP and SDSS-V are not independent.

We corrected for extinction using the \texttt{Bayestar19} 3D dust map \citep{Green2019}, as implemented in the \texttt{dustmaps} Python package \citep{Green2018}, to enable more accurate comparisons with synthetic luminosity functions and stellar populations. 
Extinction coefficients were adopted from \citet{Danielski2018}\footnote{https://www.cosmos.esa.int/web/gaia/edr3-extinction-law}. 
After applying these corrections, we calculated the extinction-corrected absolute $G$-band magnitude, \absGDered, using Gaia parallaxes.
We have also tested implementing extinction from \cite{Zhang2023} and found no significant difference in our results. 

Since main-sequence stars are faint and therefore incomplete in our sample, we include only giants for better comparison with the models. 
Giants are selected by requiring $(G_{\rm BP} - G_{\rm RP})_0 > 1$ and $(M_G)_0 < 2.8(G_{\rm BP} - G_{\rm RP})_0 - 0.5$. 
Additionally, because spectral fitting is challenging for very cool stars due to molecular absorption, we restrict our sample to stars with extinction-corrected Gaia colors $(G_{\rm BP} - G_{\rm RP})_0 < 3$, corresponding to $T_{\rm eff} \sim 3200\mathrm{K}$.
Including cooler stars does not significantly affect our results.
Finally, we exclude stars with reported [Fe/H] uncertainties greater than 0.1 in GALAH, LAMOST, and SDSS-V; stars with Gaia Renormalized Unit Weight Error (RUWE) $> 1.1$, to ensure reliable metallicities and photometry and to remove obvious binaries \citep{CastroGinard2024}; and known non-single stars in the Gaia DR3 catalog \citep{GaiaMulti2023}.

\subsection{Synthetic Stellar Populations}\label{subsec:LF}
To compare with observations, we generated synthetic stellar populations using the PARSEC v1.2S isochrones \citep{Bressan2012, Marigo2008, Girardi2002} via the CMD web interface\footnote{\url{https://stev.oapd.inaf.it/cgi-bin/cmd}}. 
We adopt a circumstellar dust composition of 60\% silicate + 40\% AlOx \citep{Groenewegen2006} and a Reimers mass-loss prescription with a scaling factor $\eta_{\rm Reimers} = 0.2$ \citep{Reimers1975}. 
Values of $\eta_{\rm Reimers}$ between 0.2 and 0.4 do not significantly alter our results.
To better match the observations, we account for the median stellar age at each metallicity, using ages from \cite{Lu2026}, and generate synthetic populations assuming a Kroupa initial mass function \citep{Kroupa2001, Kroupa2002} with a total mass of $10^6 M_{\odot}$ at each age.
This total mass was chosen to balance computational efficiency with statistical robustness, producing a sufficient number of giants while avoiding small-number statistics.

\section{Results} \label{sec:result}
\subsection{The truncated observed luminosity function}

\begin{figure*}
\includegraphics[width=0.5\textwidth]{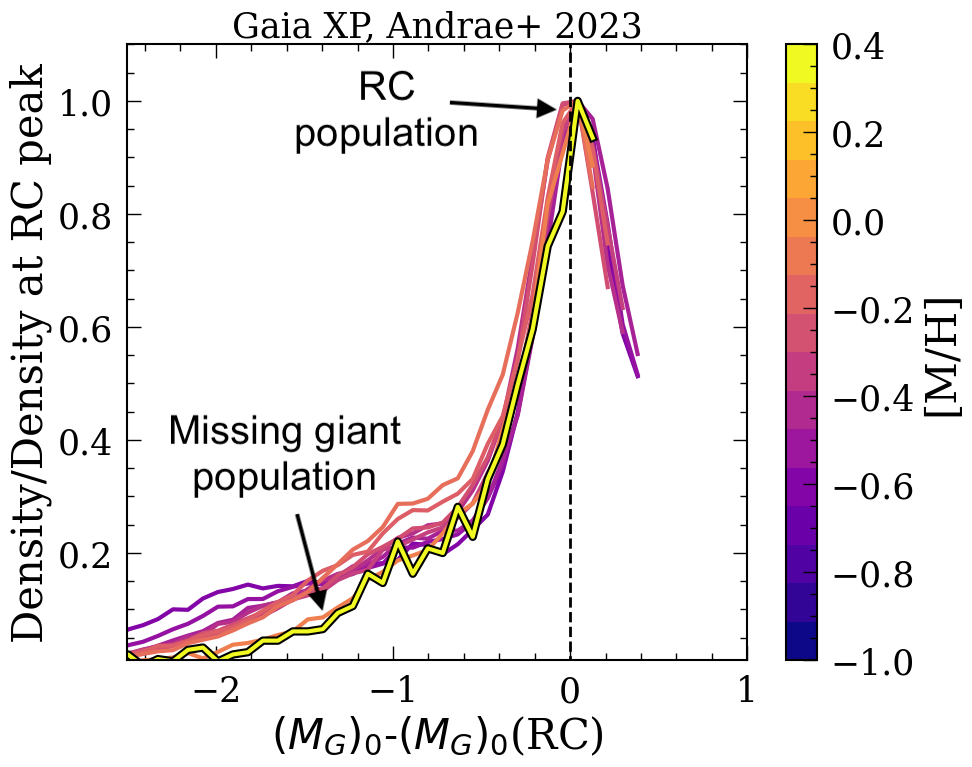}
\includegraphics[width=0.5\textwidth]{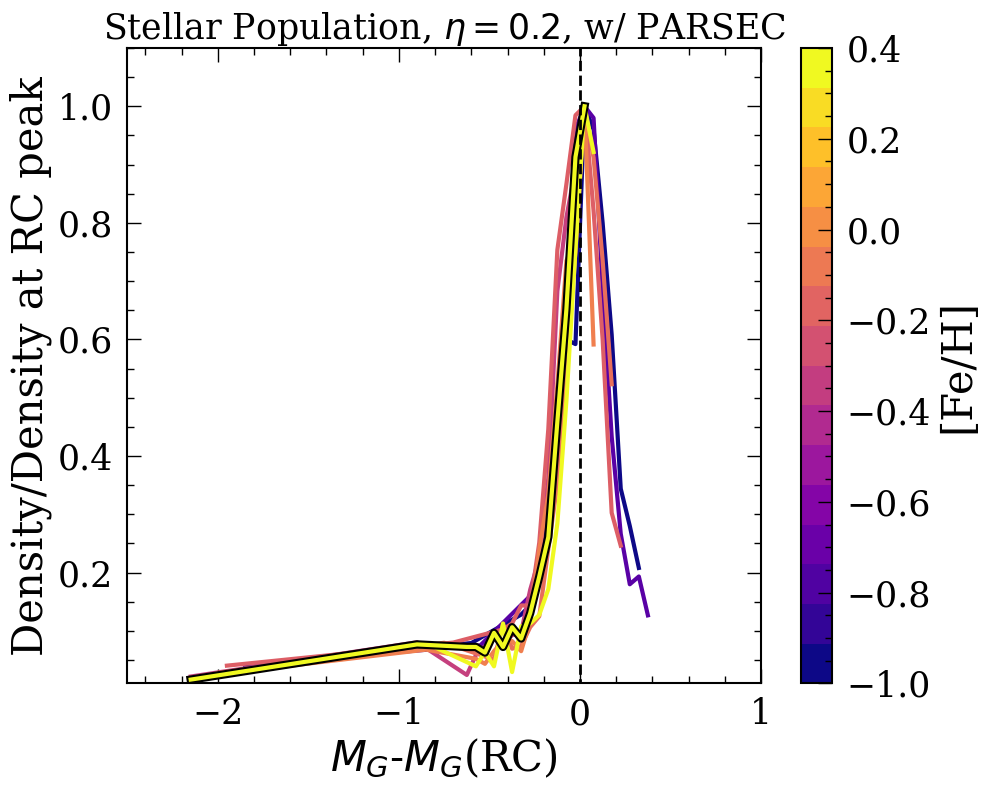}
\includegraphics[width=0.33\textwidth]{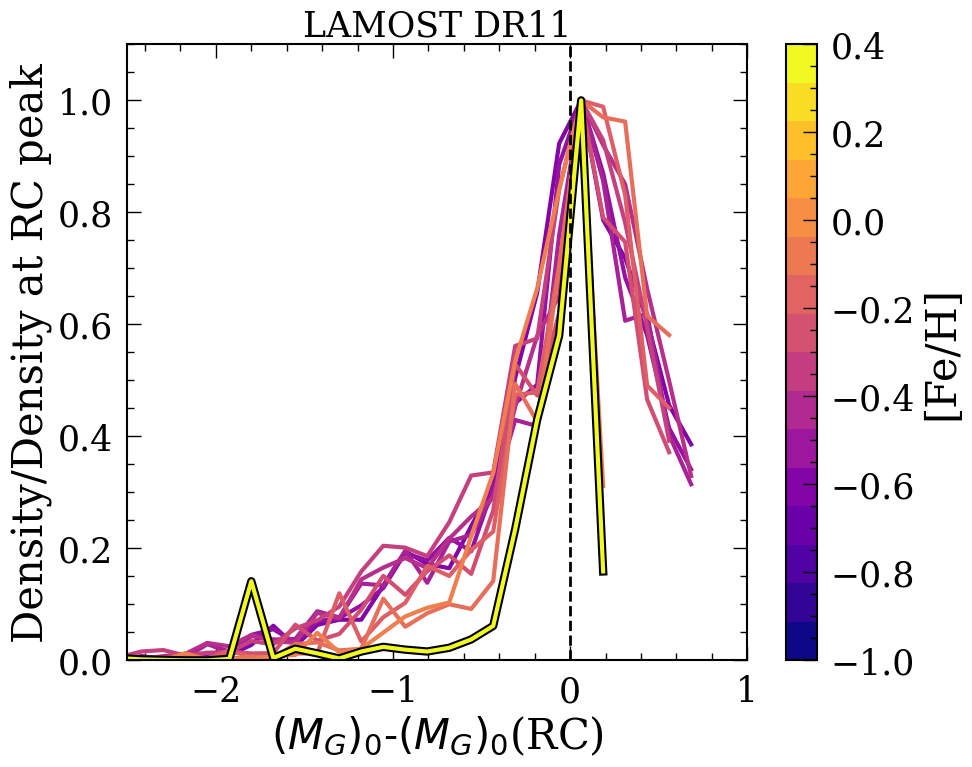}
\includegraphics[width=0.33\textwidth]{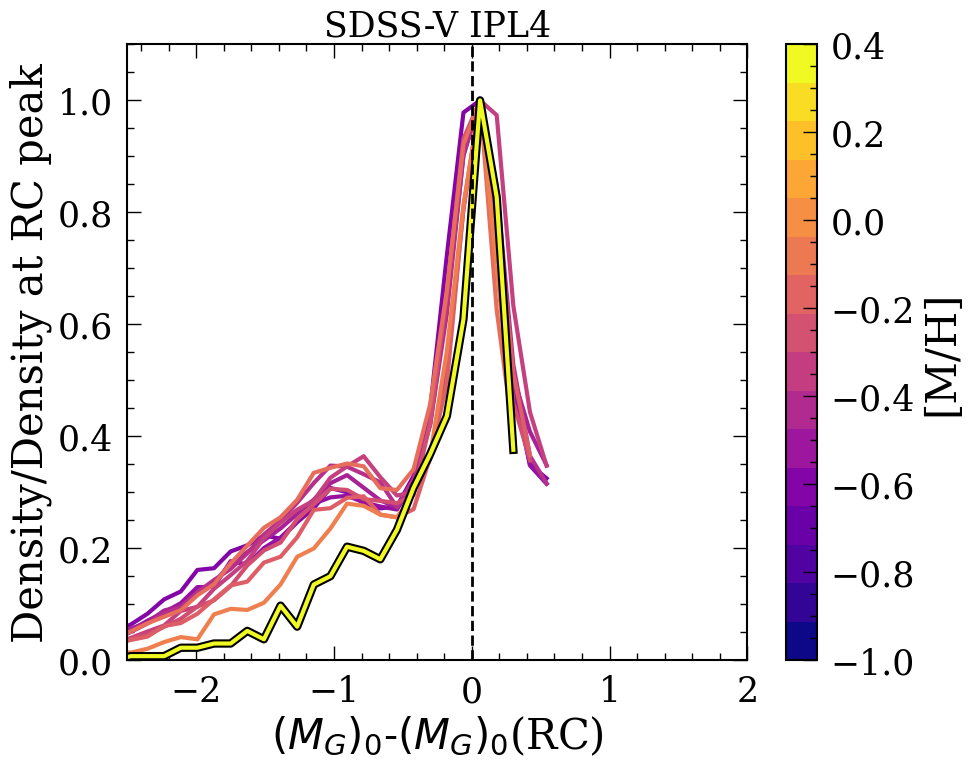}
\includegraphics[width=0.33\textwidth]{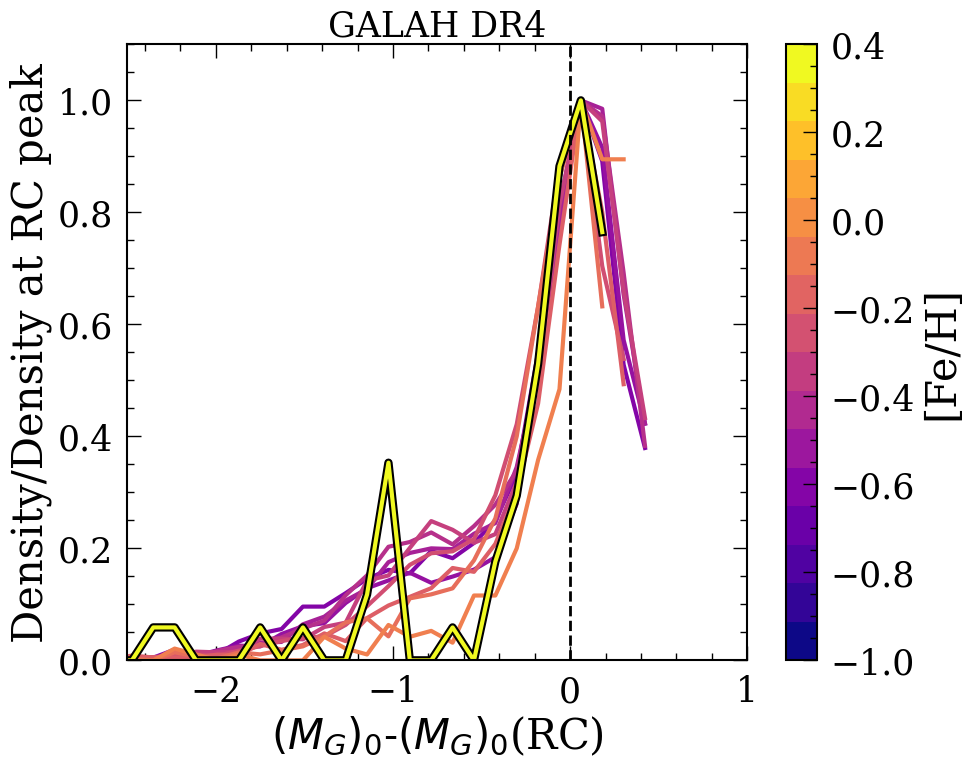}
\caption{Normalized density distribution of extinction-corrected absolute Gaia $G$-band magnitudes for Gaia XP sample (top left) and three other large-scale surveys (bottom row), as indicated in the subplot titles. 
All the histograms are also normalized to the peak of the RC distribution for better comparison.
The top right plot shows that for the synthetic population generated from PARSEC.
Only giant stars are included, selected with \absGDered $< 2.8$\bprpDered$-0.5$. 
Colors indicate metallicity, binned in 0.1 dex intervals from [Fe/H] $=-0.5$ to 0.4 for the survey data, with the last bin (yellow line with black outline) showing all stars with [Fe/H] $> 0.4$. 
Vertical dashed lines mark the red clump (\absGDered $\sim 0.5$) populations.
We shifted all the histograms to center the RC peak at 0 for better visualization. 
The truncation of the distributions beyond the red clump peak for the highest metallicity stars indicates the missing giant population in all four surveys compared to the synthetic population shown in the top right.
\label{fig:1}}
\end{figure*}

\autoref{fig:1} shows the normalized distributions of extinction-corrected absolute $G$-band magnitudes for giant stars from four spectroscopic surveys, together with a synthetic population generated using PARSEC isochrones. 
The colored curves represent stars grouped into 0.1 dex metallicity bins spanning $-0.5 \leq \mathrm{[Fe/H]} < 0.4$, with each color corresponding to a different metallicity interval.
The last bin (yellow line with black outline) includes all stars with [Fe/H] $> 0.4$ or [Fe/H] $= 0.5$ for the synthetic population. 
The black dashed lines mark the RC population at \absGDered\ $\sim 0.5$ mag.
The histograms are shifted such that the RC population is centered at zero, and the heights are normalized by the peak of the RC population to facilitate visualization.

\begin{figure}
\includegraphics[width=\columnwidth]{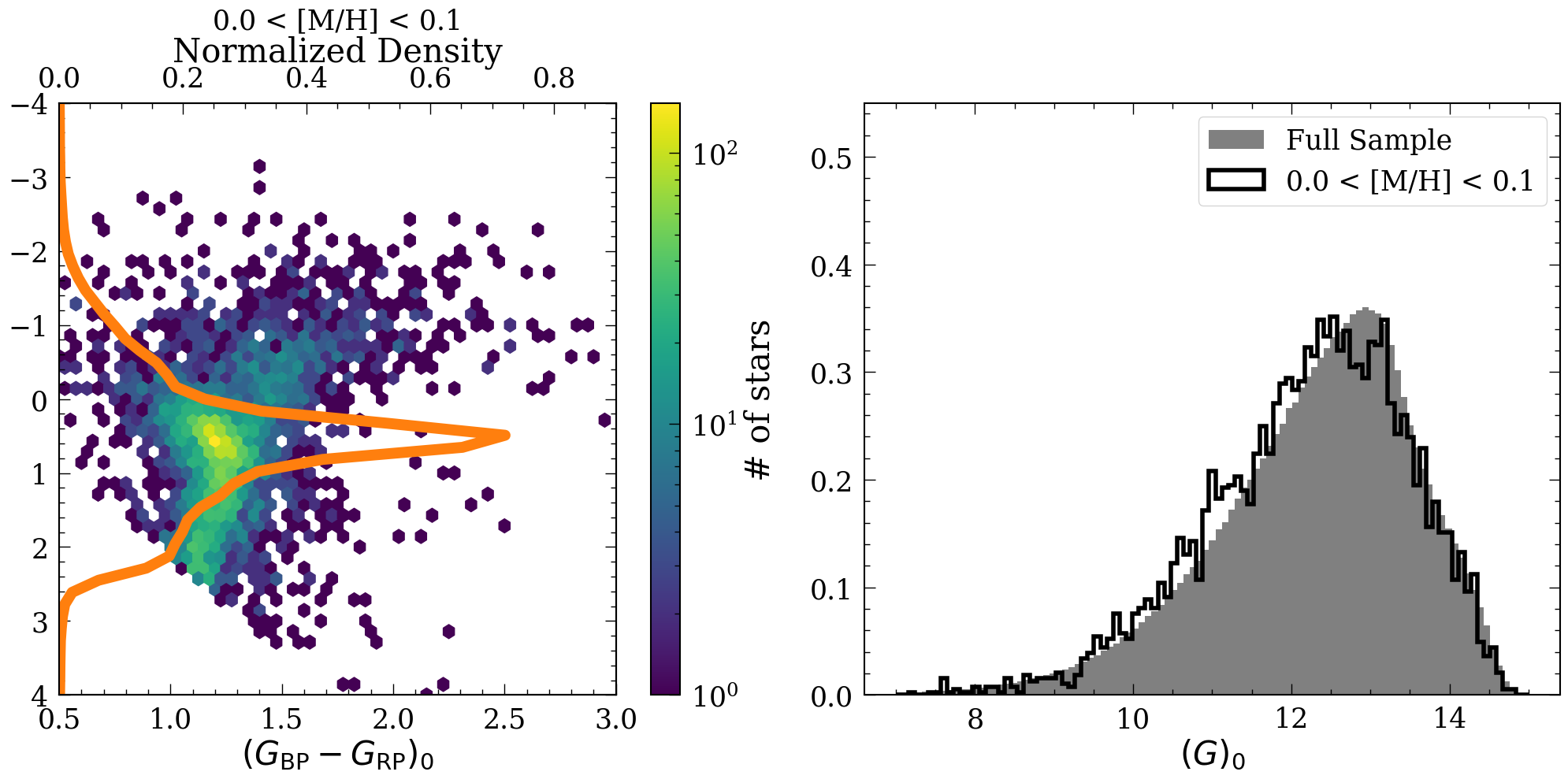}
\includegraphics[width=\columnwidth]{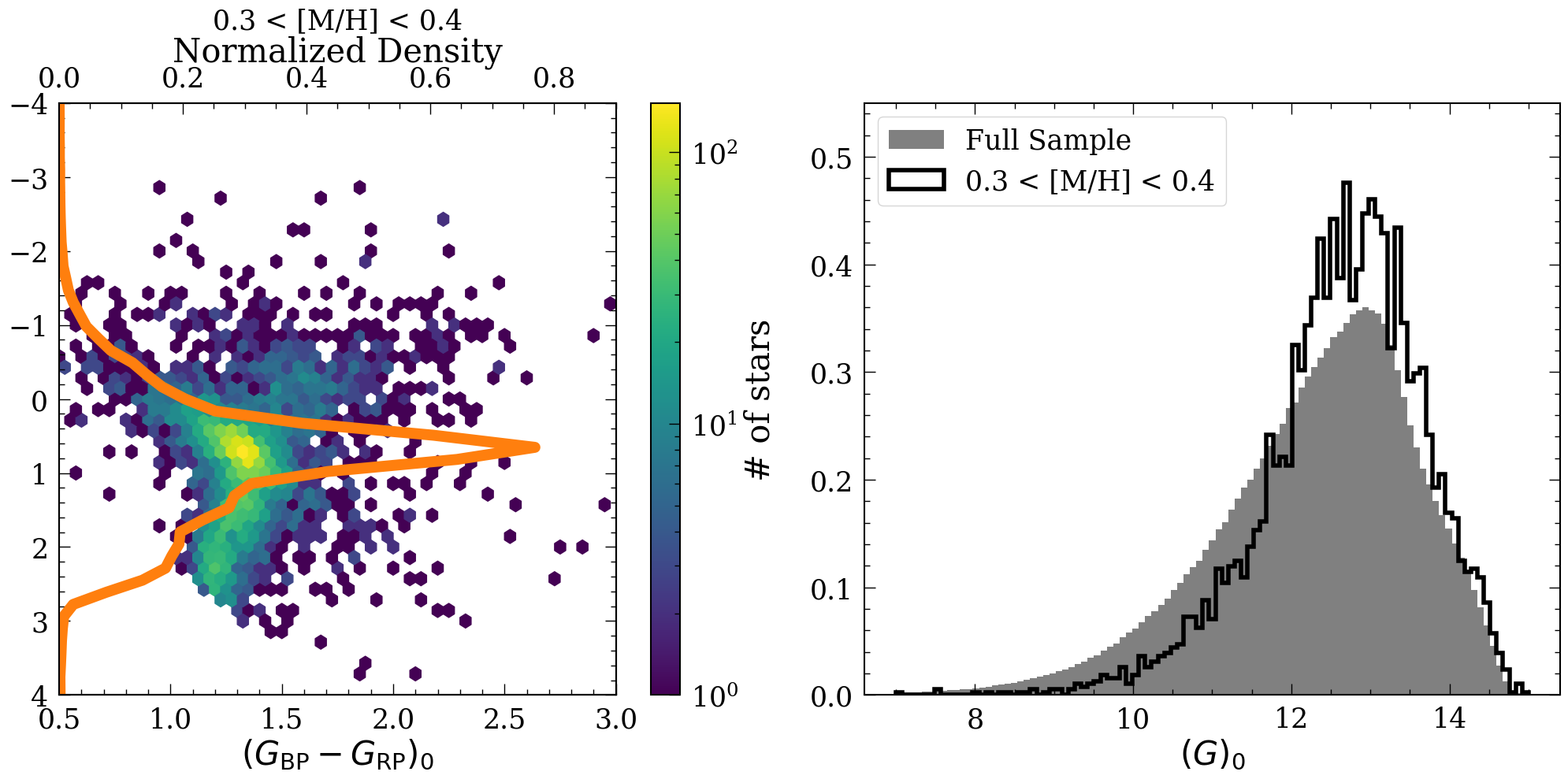}
\includegraphics[width=\columnwidth]{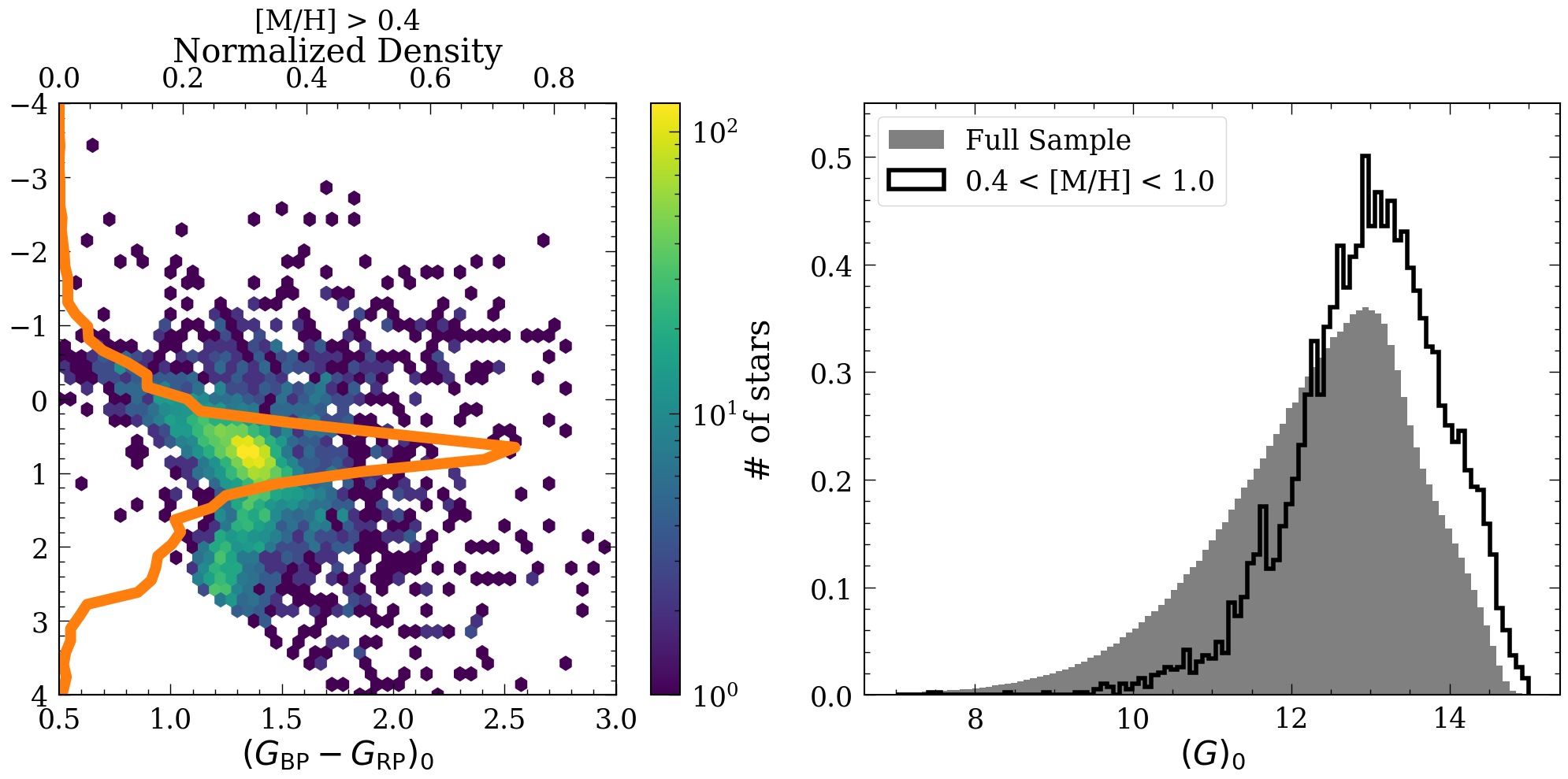}
\caption{Comparison demonstrating that the deficit of luminous giant stars is not caused by the Gaia brightness limit.
Color–magnitude diagrams (CMDs; left) and extinction-corrected apparent $G$-band magnitude ($G_0$) distributions (right) for three metallicity bins: solar (0 $<$ [M/H] $<$ 0.1, top row), metal-rich (0.3 $<$ [M/H] $<$ 0.4, middle row), and super metal-rich ([Fe/H] $>$ 0.4, bottom row). 
The orange density distributions in the CMDs correspond to \autoref{fig:1} for the same metallicity for comparison, while the gray histograms show the full $G_0$ distribution for the XGBoost sample.
At high metallicity, the truncation of the most luminous giants (\absGDered $< 0$) is apparent: the CMD shows a sharp drop in density beyond the red clump at \absGDered $\sim 0.5$, and the corresponding $G_0$ distributions for metal-rich stars lack the luminous tail, as seen on the right, suggesting the brightest stars make up the missing giant population at high metallicity. 
The peak of the $G_0$ histograms for all three metallicity bins occurs around 13 mag, indicating no significant difference in their average distances.
\label{fig:2}}
\end{figure}

Compared to the synthetic population, the giant population with \absGDered\ smaller than the RC exhibit a pronounced metallicity-dependent truncation or decrease that is absent in the synthetic population, with the deficit becoming increasingly severe toward higher metallicities.
This trend is also evident in the CMD and extinction-corrected apparent $G$-band magnitude ($G_0$) distributions, shown in \autoref{fig:2}. 
The left column displays CMD density distributions, while the right column shows $G_0$ histograms for solar-metallicity stars (top row), metal-rich stars (middle row), and super metal-rich stars (bottom row), as indicated in the subplot titles. 
The orange density distributions in the left column correspond to those in \autoref{fig:1} (top left) for the respective metallicity bins, and the full $G_0$ distribution for the XGBoost sample is shown as the gray background histogram in the right column.
Moving from solar-metallicity to super metal-rich stars, the most luminous giants progressively disappear, both in the CMD and at the bright end of the $G_0$ distribution. 
While the $G_0$ distribution for solar-metallicity stars closely resembles that of the full sample, the most metal-rich stars exhibit a more Gaussian-like distribution, lacking the bright-end tail where luminous giants would be found. 
The similar peak locations across metallicity bins suggest that the observed trend is not primarily driven by distance differences.

\subsection{Comparing with synthetic stellar population}\label{subsec:synth}
\begin{figure*}
\centering
\includegraphics[width=0.48\textwidth]{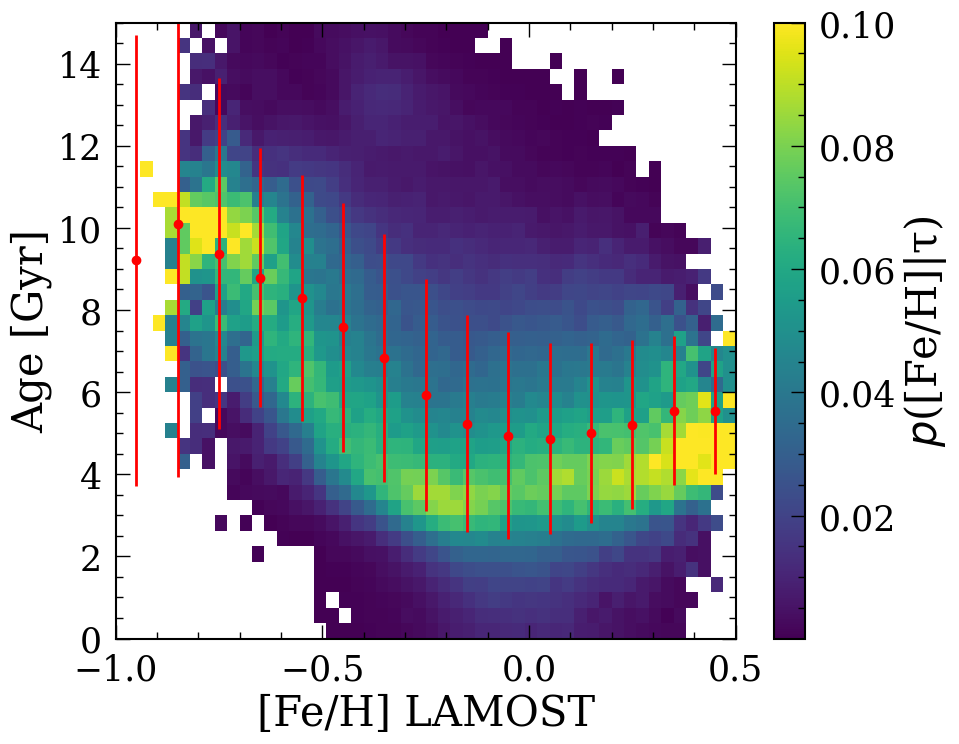}
\includegraphics[width=0.48\textwidth]{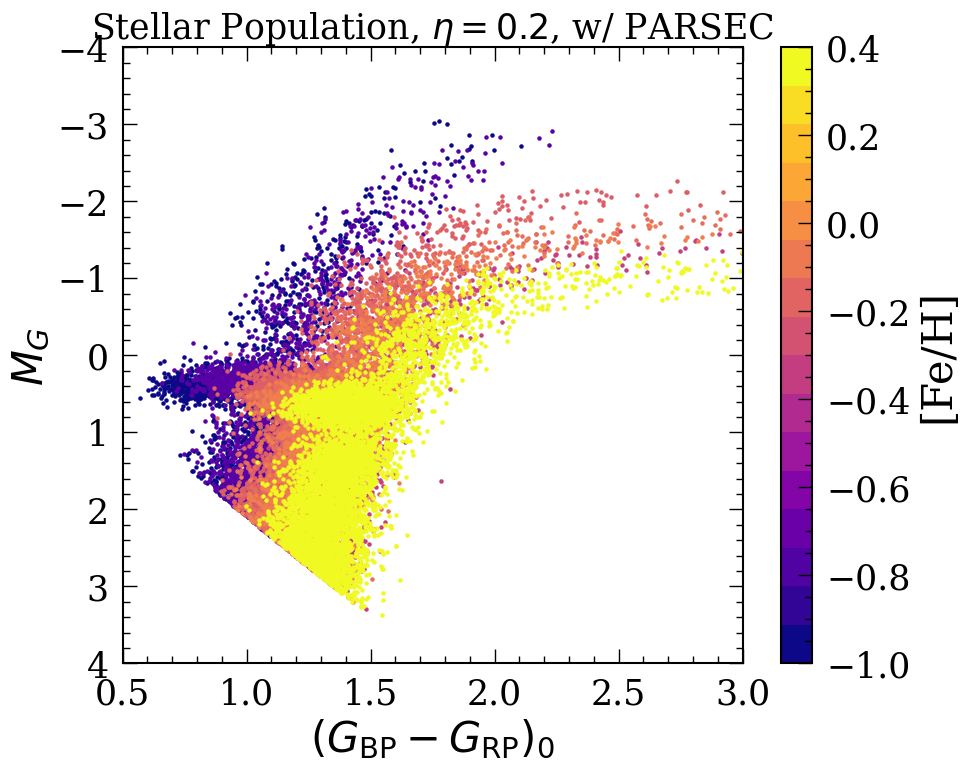}
\includegraphics[width=0.32\textwidth]{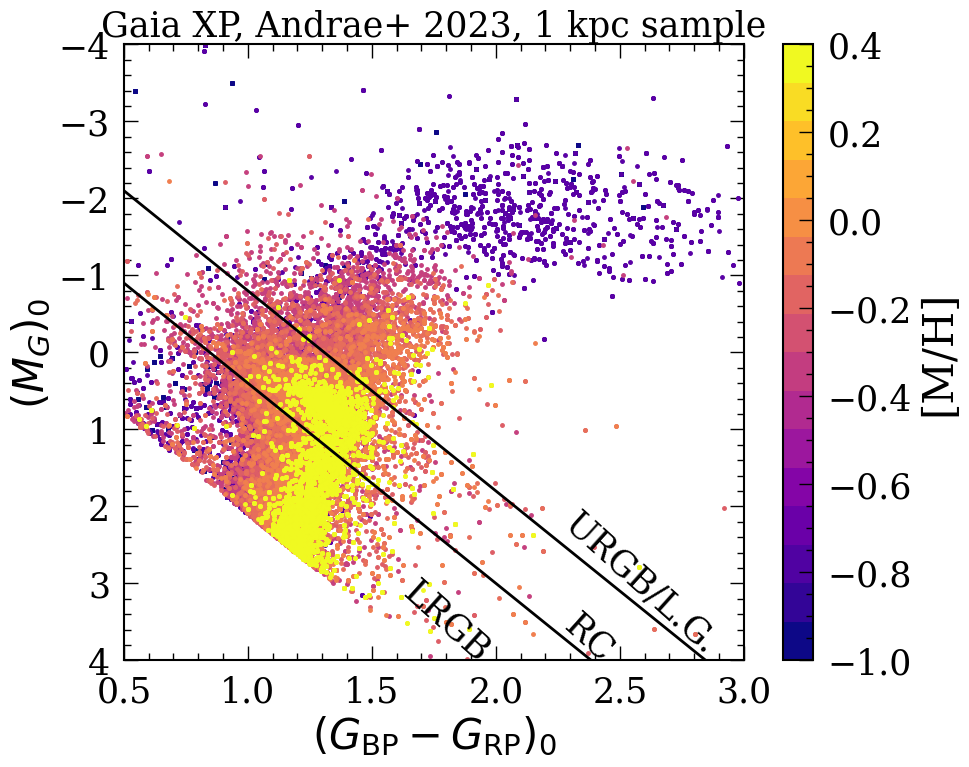}
\includegraphics[width=0.32\textwidth]{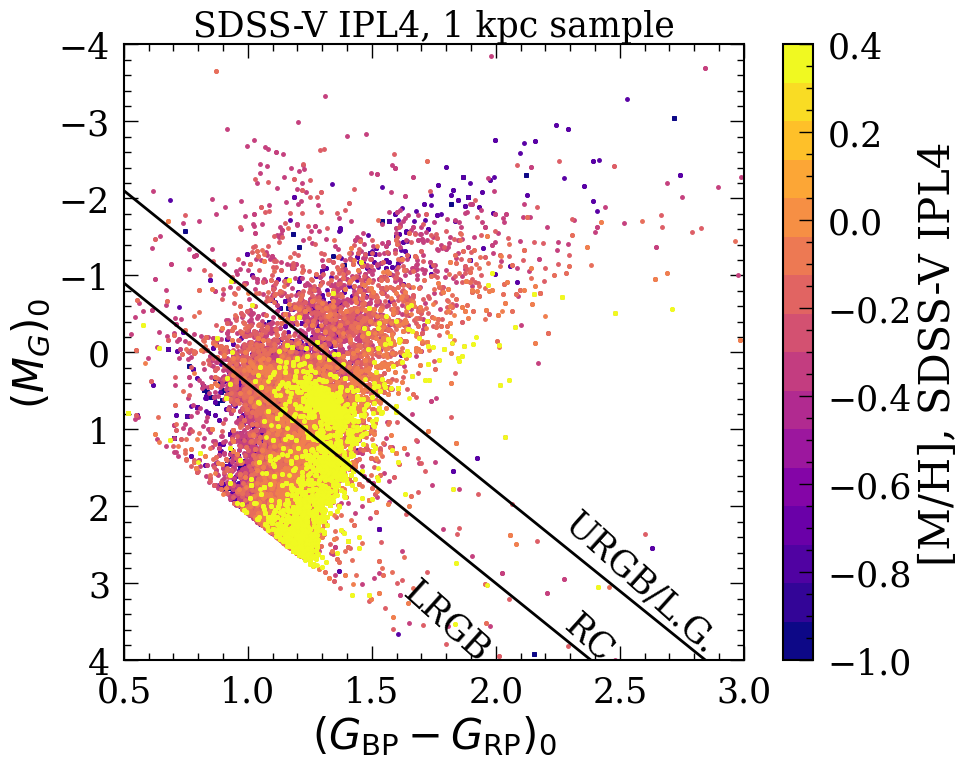}
\includegraphics[width=0.32\textwidth]{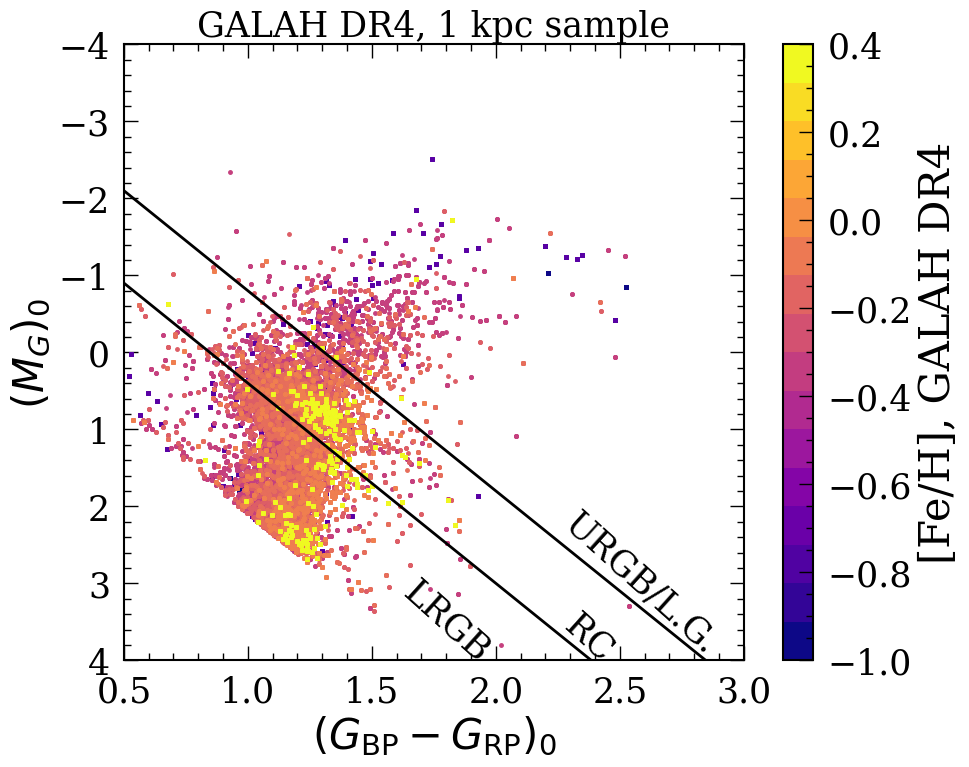}
\caption{Top left: Column-normalized age distribution as a function of metallicity, based on 1.5 million LAMOST stars from Lu \& Pinsonneault (2026). 
Selecting only the stars in common with \cite{Andrae2023} does not significantly change the resulting distributions.
Red points indicate the mean age in each metallicity bin, with error bars showing the standard deviation.
The average age is relatively constant for [Fe/H]$>-0.4$, and the distribution are similar with no additional peaks.
This suggest age cannot explain the missing giants in observations.
Top right: Synthetic CMD colored by metallicity, with Gaussian uncertainties of 0.1 mag applied to both $G_{\rm BP}-G_{\rm RP}$ and $G$ to mimic observational errors; the same giant selection criteria as in the data are applied. 
Bottom left: CMD for the Gaia XP sample within 1 kpc, with metallicity bins matched to those used in the synthetic stellar population (bottom left). 
For each metallicity bin, an equal number of stars is plotted to facilitate comparison.
The black lines define the rough separation between upper red giant branch (URGB) stars/luminous giants (L.G.), red clumps (RC), and lower red giant branch (LRGB) stars.
Bottom middle and right: same as bottom left but for SDSS-V IPL4 and GALAH DR4, respectively.
It is clear that no giant stars are missing in the highest metallicity bin compared to observation.
Unlike the data, the Synthetic stellar population does not show any missing giants at metallicity of 0.5.
\label{fig:3}}
\end{figure*}

As described in \autoref{subsec:LF}, we generated synthetic stellar populations across a range of metallicities, adopting the average age at each metallicity. 
The column-normalized age distribution is shown in the top-left panel of \autoref{fig:3}, where the red points indicate the mean age and the error bars represent the standard deviation. 
The 1.5 million stellar ages are taken from Lu \& Pinsonneault (2026), where kinematic ages are inferred from median vertical action of stars that are grouped by surface temperature, log$g$, and abundances. 
The average age remains nearly constant for [Fe/H] $> -0.4$, suggesting that population age differences are unlikely to explain the deficit of luminous giants.

The top-right panel of \autoref{fig:3} shows the synthetic CMD, color-coded by metallicity.
The bottom-left panel of \autoref{fig:3} shows the CMD for the Gaia XP sample within 1 kpc, with metallicities matched to within 0.1 dex of those in the synthetic population. 
We randomly sample 3000 stars from each metallicity bin, with replacement, to facilitate comparison.
We also convolve Gaussian uncertainties of 0.1 mag in both $G_{\rm BP}-G_{\rm RP}$ color and $G$-band magnitude for the synthetic population. 
The same selection criteria as in the observational data are then applied to isolate giant stars.
The transformation from isochrone parameters to Gaia photometry follows \citet{Riello2021}. 
The bottom middle and right panel shows the same but for SDSS-V and GALAH DR4, respectively.
It is apparent that, compared to the synthetic isochrones, the relative number of luminous giants decreases with increasing metallicity in the three surveys.

\subsection{Fraction of luminous giants, red clumps, and upper/lower red giant branch stars}\label{subsec:ratio}
\begin{figure*}
\includegraphics[width=\textwidth]{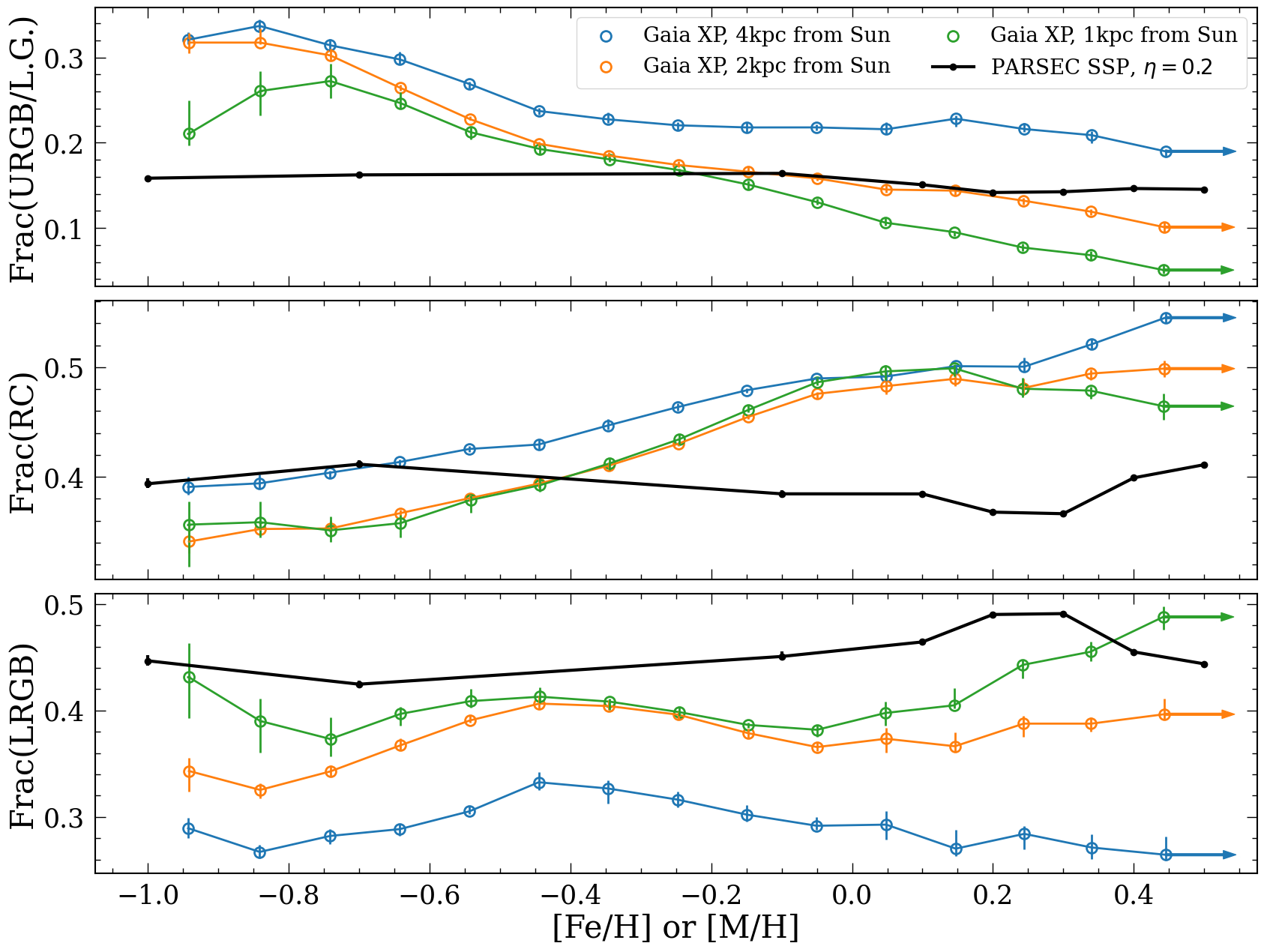}
\caption{Fraction of upper red giant branch/luminous giant (URGB/L.G.; top), red clump (RC; middle), and lower red giant branch (LRGB; bottom) stars compared to the total population as a function of metallicity. 
Results are shown for the Gaia XP sample within 4, 2, and 1 kpc of the Sun, as well as for the synthetic populations, as indicated in the legend. 
The final metallicity bin for the Gaia sample includes all stars with [Fe/H] $> 0.4$, denoted by right-pointing arrows.
The trends are similar regardless of the distance selection, meaning distance is not the main cause of the missing giants.
The trend is shallower moving towards larger distances, likely reflect the incompleteness.
However, the decrease in the fraction of URGB/L.G is most apparent within 1 kpc of the Sun, where almost no URGB/L.G exist despite the existence of RC and LGB population.
This is in clear disagreement with the synthetic population, where all fractions remain approximately constant.
If this effect is physical in origin, it likely impacts only a small fraction of the stellar population, as the fraction of RC stars does not decrease in the same way as the URGB and luminous giant populations.
\label{fig:4}}
\end{figure*}

Because the tip of the red giant branch shifts toward higher \absGDered\ with increasing metallicity (as shown for the synthetic population in the bottom-right panel of \autoref{fig:3}), we directly compare the relative fractions of stars in different evolutionary stages to determine whether luminous giants are preferentially missing at the highest metallicities. 
If these extremely metal-rich stars experience a stochastic episode of enhanced mass loss, we would expect the fractions of upper red giant branch (URGB) stars (i.e., luminous giants or L.G.) and red clump (RC) stars to decrease relative to the fraction of lower red giant branch (LRGB) stars.
We combine the URGB and L.G. populations because both consist of RGB stars that have not yet evolved into the RC phase.
We computed the fraction of URGB/L.G., LRGB, and RC populations using the definition shown in \autoref{fig:3} (top right panel). The boundaries separating URGB/L.G., RC, and LRGB were determined independently within each metallicity bin by vertically shifting the relation \absGDered = 2.8 \bprpDered\ to trace the red clump locus.
The intercept offset defining the red clump region was fixed to 1 mag to ensure consistency across all metallicity bins.
This calculation was performed for the Gaia XP sample for stars within 4, 2, and 1 kpc of the Sun, as indicated in the legend. 
We adopt the Gaia XP sample, as it provides the largest set of metallicity measurements with minimal selection bias. 
For the synthetic populations, the URGB/L.G., RC, and LRGB boundaries are defined analogously to the observations, using horizontal cuts in \absGDered, with the RC width fixed at 0.6 mag.
For both the observations and the synthetic populations, the LRGB cutoff is set 1.5 mag above the LRGB–RC boundary to facilitate better comparison.
Uncertainties are estimated by perturbing the metallicity, \bprpDered, and \absGDered by 0.1, as well as shifting the boundaries between RC, LRGB, and URGB by 0.1 mag in \absGDered, and computing the 16$^{\rm th}$, 50$^{\rm th}$, and 84$^{\rm th}$ percentiles of the resulting posterior distribution.
Only metallicity bins containing more than 900 stars are shown, ensuring a typical bin width of $\sim$0.5 mag in \absGDered.
\autoref{fig:4} shows the results. 

The general trends in the Gaia XP data are consistent across different distance selections, suggesting they are unlikely to be driven solely by selection effects. 
The trends appear weaker as the sample includes stars at larger distances, indicating that some populations are likely missing from the more distant samples.
However, the most complete sample ($<$ 1 kpc), shown in green, exhibits the strongest trend, despite an increase in the LRGB fraction, the URGB/L.G. fraction decreases, pointing to a genuinely missing population. 
The fraction of RC stars increases with metallicity, likely driven by the corresponding decrease in the URGB/L.G. fraction. 
This behavior is in strong disagreement with the synthetic population, where all fractions remain approximately constant.
If this effect is physical in origin, it likely impacts only a small fraction of the stellar population.

\section{On the Missing Giants}\label{sec:invest}
In this section, we investigate several possible non-astrophysical pathways that could explain the observed deficit of metal-rich giants. 
In a subsequent paper, we perform an asteroseismic analysis (Howell et al. in prep.) to test whether extreme mass loss could contribute to this population deficit by measuring asteroseismic masses for stars with Gaia XP ([M/H] $>$ 0.45). 
We find that the average RGB stellar mass is consistent with expectations, while RC stars exhibit moderately lower masses, providing tentative evidence for enhanced mass loss.

\subsection{Bias in detecting metal-rich stars}\label{subsec:bias}
One possibility is that these extreme metal-rich stars are preferentially located at larger distances and are therefore more difficult to detect, particularly at the luminous end of the giant branch. 
In this scenario, selection effects could artificially suppress the number of bright giants in the highest metallicity bins.
However, several lines of evidence argue against this explanation. 
First, the $G_0$ distributions for different metallicity bins peak at similar magnitudes, indicating comparable distance distributions and suggesting that the absence of luminous giants is not driven by distance effects. 
Moreover, the missing giants occupy the most luminous end of the distribution, which is clearly absent in the bottom panel of \autoref{fig:1} when comparing the most metal-rich $G_0$ distribution to that of the full sample.
In addition, the missing giant trends with metallicity persists when restricting the sample to progressively smaller volumes (4, 2, and 1 kpc), where completeness is expected to be high and selection biases minimized.
Finally, photometric uncertainties and extinction corrections could, in principle, scatter stars out of the luminous giant region. 
Nevertheless, our analysis accounts for these effects, and the extinction corrections appear robust, as the RC population peaks at the same \absGDered, consistent with expectations from the synthetic population.
Taken together, these considerations indicate that observational biases alone are unlikely to explain the missing luminous giants at high metallicity.

\subsection{Misclassifications of the metal-rich stars}\label{subsec:misclass}
Another potential nonphysical cause could arise from spectroscopic limitations. 
Metal-rich stars exhibit stronger line blanketing, which can complicate continuum normalization and spectral fitting, potentially leading to systematic errors in metallicity estimates or incompleteness at high [Fe/H].
This is especially difficult in low resolution spectra such as those from Gaia XP.
However, when comparing with the overlapping samples from GALAH, LAMOST, and SDSS-V IPL4, we find excellent agreement across the full range of \absGDered\ from [Fe/H] $=-1$ to $0.5$, consistent with \citet{Andrae2023}.
If line blanketing were the primary cause of the missing giants, its impact should be reduced when using the inferred stellar parameters from SDSS-V. 
However, the deficit of luminous giants remains equally apparent in the SDSS-V IPL-4 data (\autoref{fig:3}), arguing against line blanketing as the primary explanation.

\begin{figure}
\includegraphics[width=\columnwidth]{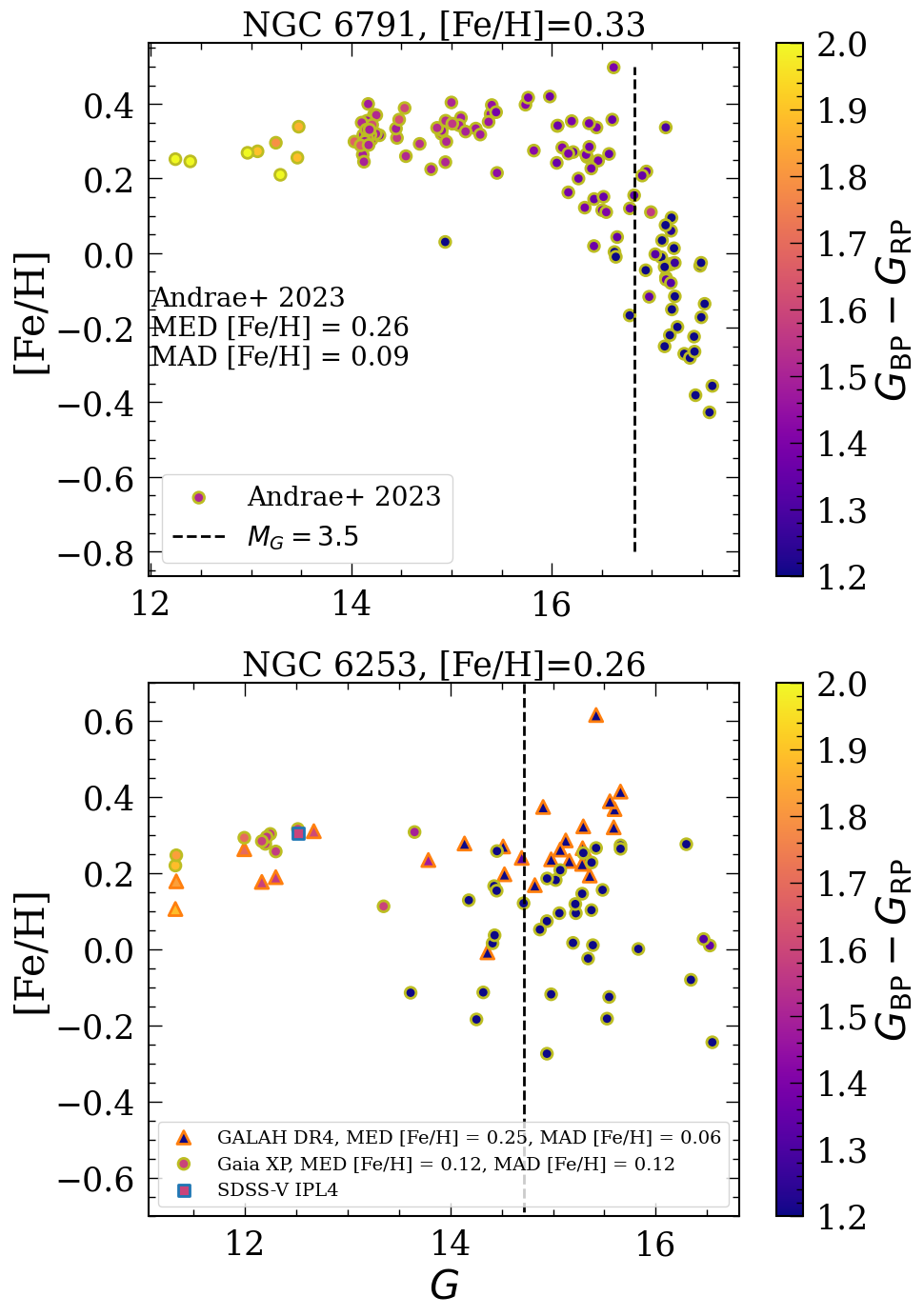}
\caption{Metallicity reported from various surveys as a function of $G$ magnitude for member stars of the metal-rich clusters NGC 6791 and NGC 6253. 
The colored outlines and marker shapes indicate the survey of origin, as shown in the legend, while the colors represent $G_{\rm BP}-G_{\rm RP}$. 
The vertical dashed line marks the subgiant population. 
Cluster memberships are taken from \citet{Hunt2024}, with a membership probability threshold of $>70$\%. 
It is clear that, within the parameter space used in this work ($G < 16$, \absGDered $< 3.5$, and $0.5 <$\bprpDered$ < 3$), the metallicity distributions are broadly consistent with cluster values, suggesting that Gaia XP spectra for metal-rich giants are relatively reliable.
\label{fig:5}}
\end{figure}

However, it is possible that, despite the agreement between surveys, the metallicities are systematically biased in a similar way across all datasets \citep[e.g.,][]{Saad2026}.
To test this, we compare the metallicities in the surveys reported for the members of the metal-rich cluster NGC 6791 \citep[{[Fe/H]} $\sim 0.33$,][]{Perren2023} and NGC 6253 \citep[{[Fe/H]} $\sim 0.26$,][]{Perren2023}.
Cluster memberships are taken from \citet{Hunt2024}, requiring a membership probability greater than 70\%.
\autoref{fig:5} shows the metallicity reported from various surveys as a function of $G$ magnitude.
The points are color-coded by $G_{\rm BP}-G_{\rm RP}$, and the vertical dashed line marks the subgiant population.
Within the parameter space used in this work ($G < 16$, \absGDered $< 3.5$, and $0.5 <$\bprpDered$ < 3$), the metallicity distributions are broadly consistent with cluster values, suggesting that Gaia XP spectra for metal-rich giants are relatively reliable.
A slight trend with magnitude is seen in NGC 6791, and correcting for this trend does not significantly affect our results.

\begin{figure}
\includegraphics[width=\columnwidth]{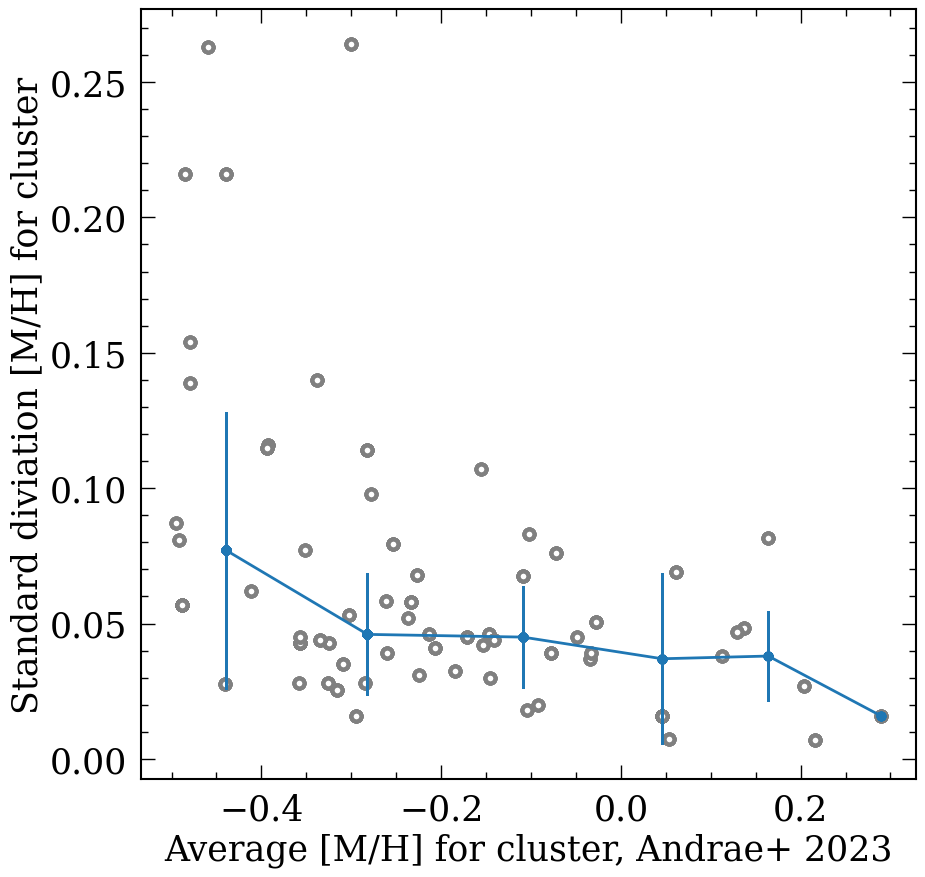}
\caption{The average and standard deviation of the [M/H] measurements from \cite{Andrae2023} for URGB/L.G. stars in clusters with membership probability $>0.8$ from \cite{Hunt2024}. 
Only clusters containing more than 5 URGB/L.G. stars are shown. 
The intrinsic [M/H] scatter within each cluster is small ($\sim$0.05 dex) and remains approximately constant for clusters with [M/H] $>-0.3$, despite the declining number of URGB/L.G. stars with metallicity as shown in \autoref{fig:4}.
\label{fig:ocfeh}}
\end{figure}

To further validate this result, we cross-matched the full cluster sample from \citet{Hunt2024} with the metallicity measurements from \cite{Andrae2023} for stars in the URGB/L.G. region (defined in the bottom-left panel of \autoref{fig:3}). 
We find that the cluster metallicity scatter remains small ($\sim$0.05 dex), and the median scatter is nearly constant for clusters with [M/H] $>-0.3$, despite the declining fraction of URGB/L.G. stars (see \autoref{fig:ocfeh}).

In addition, we test the metallicity consistency between surveys at the metal-rich end. 
We select stars with reported metallicities of [Fe/H] $>$ 0.3 dex in both Gaia XP and the comparison surveys. 
This yields 1,173 stars matched to SDSS-V, 602 matched to GALAH DR4, and 30,448 matched to LAMOST DR11. 
The mean residuals (Gaia XP minus survey) and standard deviations are 0.00$\pm$0.04 dex, 0.00$\pm$0.07 dex, and $-$0.05$\pm$0.08 dex for SDSS-V, GALAH DR4, and LAMOST DR11, respectively, indicating excellent agreement among the surveys.

However, we cannot rule out the possibility that some of the missing extremely metal-rich giants are misclassified as more metal-poor stars, as the available sample is not sufficient to robustly test this scenario.
If this is true, using large-scale survey data such as fitting chemical evolution models need to take this effect into account.

\section{Conclusion}\label{sec:conclude}
We have presented evidence for a systematic deficit of luminous red giant branch stars at metallicity $>$ 0.4 using Gaia XP-based stellar parameters combined with higher resolution spectroscopic surveys (LAMOST, SDSS-V, and GALAH). 
Across all datasets, we find that while the red clump and lower red giant branch populations remain relatively stable, the most luminous giants progressively disappear with increasing metallicity.
This trend is robust to distance cuts (1–4 kpc), extinction corrections, and quality selections, indicating that it is unlikely to be caused by simple observational selection effects.
synthetic stellar populations constructed using PARSEC isochrones, realistic ages, and observational uncertainties reproduce the general shape of the observed magnitude distributions but do not predict a decline in the luminous giant fraction with metallicity.
This discrepancy suggests that standard stellar population models do not capture the observed behavior. 
However, the relatively constant red clump population across metallicity suggests that, if the effect is physical, the underlying process may be stochastic and can only affect a small fraction of stars. 
If robust, this result could have important implications for HeWD formation, stellar physics at the highest metallicities, and the initial mass function in massive metal-rich elliptical galaxies.

We investigated several possible explanations, including non-physical effects such as selection biases and unreliable metallicity determinations, and find no evidence that either can account for the observed trend. 
While the metallicity estimates themselves may be uncertain for extremely cool and faint objects, this is unlikely to affect the region of interest. 

\begin{acknowledgments}
We thank K. Z. Stanek for originally suggesting this idea and the title.
We thank Christopher Kochanek, Dominick Rowan, Jamie Tayar, Yaguang Li, and Tim Bedding for providing helpful feedback to the paper.
MHP acknowledges support from NASA grant 80NSSC24K0637 in the revision stage.
J.G.F-T gratefully acknowledges the support provided by ANID Fondecyt Regular No. 1260371.
This work has made use of data from the European Space Agency (ESA) mission Gaia,\footnote{\url{https://www.cosmos.esa.int/gaia}} processed by the Gaia Data Processing and Analysis Consortium (DPAC).\footnote{\url{https://www.cosmos.esa.int/web/gaia/dpac/consortium}} 
Funding for the DPAC has been provided by national institutions, in particular the institutions participating in the Gaia Multilateral Agreement.
This research also made use of public auxiliary data provided by ESA/Gaia/DPAC/CU5 and prepared by Carine Babusiaux. 
% SIMBAD, Vizier, ADS
This research has also made use of NASA's Astrophysics Data System.
% SDSS V
Funding for the Sloan Digital Sky Survey V has been provided by the Alfred P. Sloan Foundation, the Heising-Simons Foundation, the National Science Foundation, and the Participating Institutions. SDSS acknowledges support and resources from the Center for High-Performance Computing at the University of Utah. SDSS telescopes are located at Apache Point Observatory, funded by the Astrophysical Research Consortium and operated by New Mexico State University, and at Las Campanas Observatory, operated by the Carnegie Institution for Science. The SDSS web site is \url{www.sdss.org}.

SDSS is managed by the Astrophysical Research Consortium for the Participating Institutions of the SDSS Collaboration, including the Carnegie Institution for Science, Chilean National Time Allocation Committee (CNTAC) ratified researchers, Caltech, the Gotham Participation Group, Harvard University, Heidelberg University, The Flatiron Institute, The Johns Hopkins University, L'Ecole polytechnique f\'{e}d\'{e}rale de Lausanne (EPFL), Leibniz-Institut f\"{u}r Astrophysik Potsdam (AIP), Max-Planck-Institut f\"{u}r Astronomie (MPIA Heidelberg), Max-Planck-Institut f\"{u}r Extraterrestrische Physik (MPE), Nanjing University, National Astronomical Observatories of China (NAOC), New Mexico State University, The Ohio State University, Pennsylvania State University, Smithsonian Astrophysical Observatory, Space Telescope Science Institute (STScI), the Stellar Astrophysics Participation Group, Universidad Nacional Aut\'{o}noma de M\'{e}xico, University of Arizona, University of Colorado Boulder, University of Illinois at Urbana-Champaign, University of Toronto, University of Utah, University of Virginia, Yale University, and Yunnan University.

\end{acknowledgments}


\vspace{5mm}
\facilities{Gaia, Sloan \citep{Gunn2006}, Du Pont \citep{Bowen1973}, AAT, LAMOST}


\software{\texttt{astropy} \citep{astropy:2013, astropy:2018, astropy2022}, \texttt{Matplotlib} \citep{matplotlib}, \texttt{NumPy} \citep{Numpy}, \texttt{Pandas} \citep{pandas}, \texttt{SciPy} \citep{SciPy}, \texttt{ChatGPT} \citep{openai2025chatgpt}, \texttt{dustmaps} \citep{Green2018}}



%\appendix

%\section{Appendix information}


\bibliography{sample631}{}
\bibliographystyle{aasjournal}



\end{document}